\documentclass[apjl]{emulateapj}
\usepackage{graphicx,epsfig,epsf,rotating,fancyhdr,amsmath,natbib}
\usepackage{apjfonts}

\newcommand{\kms}{km~s$^{-1}$}
\newcommand{\app}{{$\approx$}}
\newcommand{\arc}{{\arcsec}}

\shorttitle{FINE-SCALE STRUCTURE IN CORONAL LOOPS}
\shortauthors{Scullion et al.}

\begin{document}

\title{UNRESOLVED FINE-SCALE STRUCTURE IN SOLAR CORONAL LOOP-TOPS}

\author{E.~Scullion\altaffilmark{1,2}, L.~Rouppe van der Voort\altaffilmark{1}, S.~Wedemeyer\altaffilmark{1}, P.~Antolin\altaffilmark{3}  }

\altaffiltext{1}{Institute of Theoretical Astrophysics, University of Oslo, P.O. Box 1029, Blindern, NO-0315 Oslo, Norway}
\altaffiltext{2}{Trinity College Dublin, College Green, Dublin 2, Ireland: scullie@tcd.ie}
\altaffiltext{3}{National Astronomical Observatory of Japan, 2-21-1 Osawa, Mitaka, Tokyo 181-8588, Japan}

\begin{abstract} 

New and advanced space-based observing facilities continue to lower the resolution limit and detect solar coronal loops in greater detail. We continue to discover even finer sub-structures within coronal loop cross-sections, in order to understand the nature of the solar corona. Here, we push this lower limit further to search for the finest coronal loop sub-structures, through taking advantage of the resolving power of the Swedish 1-m Solar Telescope (SST) / CRisp Imaging Spectro-Polarimeter (CRISP), together with co-observations from the Solar Dynamics Observatory (SDO) / Atmospheric Image Assembly (AIA). High resolution imaging of the chromospheric H$\alpha$~656.28~nm spectral line core and wings can, under certain circumstances, allow one to deduce the topology of the local magnetic environment of the solar atmosphere where its observed. Here, we study post-flare coronal loops, which become filled with evaporated chromosphere that rapidly condenses into chromospheric clumps of plasma (detectable in H$\alpha$) known as a coronal rain, to investigate their fine-scale structure. We identify, through analysis of three datasets, large-scale catastrophic cooling in coronal loop-tops and the existence of multi-thermal, multi-stranded sub-structures. Many cool strands even extend fully-intact from loop-top to foot-point. We discover that coronal loop fine-scale strands can appear bunched with as many as 8 parallel strands, within an AIA coronal loop cross-section. The strand number density vs cross-sectional width distribution, as detected by CRISP within AIA-defined coronal loops, most-likely peaks at well below 100~km and currently 69\% of the sub-structure strands are statistically unresolved in AIA coronal loops. 

\end{abstract}


\section{INTRODUCTION}

\par It has been long claimed \citep[for e.g.,][and earlier]{1993ApJ...405..767G} that coronal loops consist of bundles of thin strands, to scales below the current instrumental resolution. Today,  that statement continues to remain as prevalent as ever. Coronal loops were first detected in coronagraphic observations in the 1940s \citep{1991plsc.book.....B}. These loops are observed to extend into the low plasma-$\beta$ environment of the solar corona, arching over active regions, and are filled with relatively dense plasma (in the range of $\sim$10$^8$-10$^{10}$ cm$^{-3}$) and confined by a dipole-like magnetic field \citep[and references therein]{2004psci.book.....A, 2010LRSP....7....5R}. Coronal loops can have different temperatures and are observed in Extreme Ultra-Violet (EUV) from $\sim$10$^5$~K (cool loops) to a few 10$^6$~K (warm loops) and up to a few 10$^7$~K (flaring loops). In coronal loops, neighboring field lines are considered to be thermally isolated, hence, each field line can be considered independently which we call a "strand" herein. 

\par To explain the nature of coronal loops is to understand the origin of solar coronal heating. One fundamental issue is that we do not know the spatial scale of the coronal heating mechanism \citep{2010LRSP....7....5R}. It has been considered that in order to form stable over-dense, warm coronal loops it may be required to assume that coronal loops consist of unresolved magnetic strands, each heated impulsively, non-uniformly and sequentially \citep{1993ApJ...405..767G, 1994ApJ...422..381C, 2000SoPh..193...53K, 2001ApJ...553..440K, 2005ApJ...633..489R, 2006SoPh..234...41K, 2008ApJ...682.1351K}. At a typical spatial resolution (of most current space-based instruments observing from EUV to higher energies) of $\sim$1000~km ($\sim$1.5\arc: 1 arc second (\arc) \app720~km) it is likely that most observations represent superpositions of hundreds of unresolved strands at various stages of heating and cooling \citep{2006SoPh..234...41K}. Other studies based both on models and on analysis of observations independently suggest that elementary loop components should be even finer, with typical cross-sections of the strands on the order of 10-100~km \citep{2003SoPh..216...27B, 2004ApJ...605..911C, 2007ApJ...661..532D}. The space-based \citep[{\it Hinode}:][]{2007SoPh..243....3K} EUV Imaging Spectrometer \citep[EIS: ][]{2007SoPh..243...19C} has been used together with the Solar Dynamics Observatory (SDO) / Atmospheric Image Assembly \citep[AIA: ][]{2012SoPh..275...17L} to investigate the fundamental spatial scales of coronal loops and the results suggest that most coronal loops remain unresolved \citep{2012AAS...22030903U}, given the 864~km and 1440~km resolution of SDO/AIA and Hinode / EIS, respectively.


\begin{figure*}[!ht]
\includegraphics[scale=0.58, angle=0]{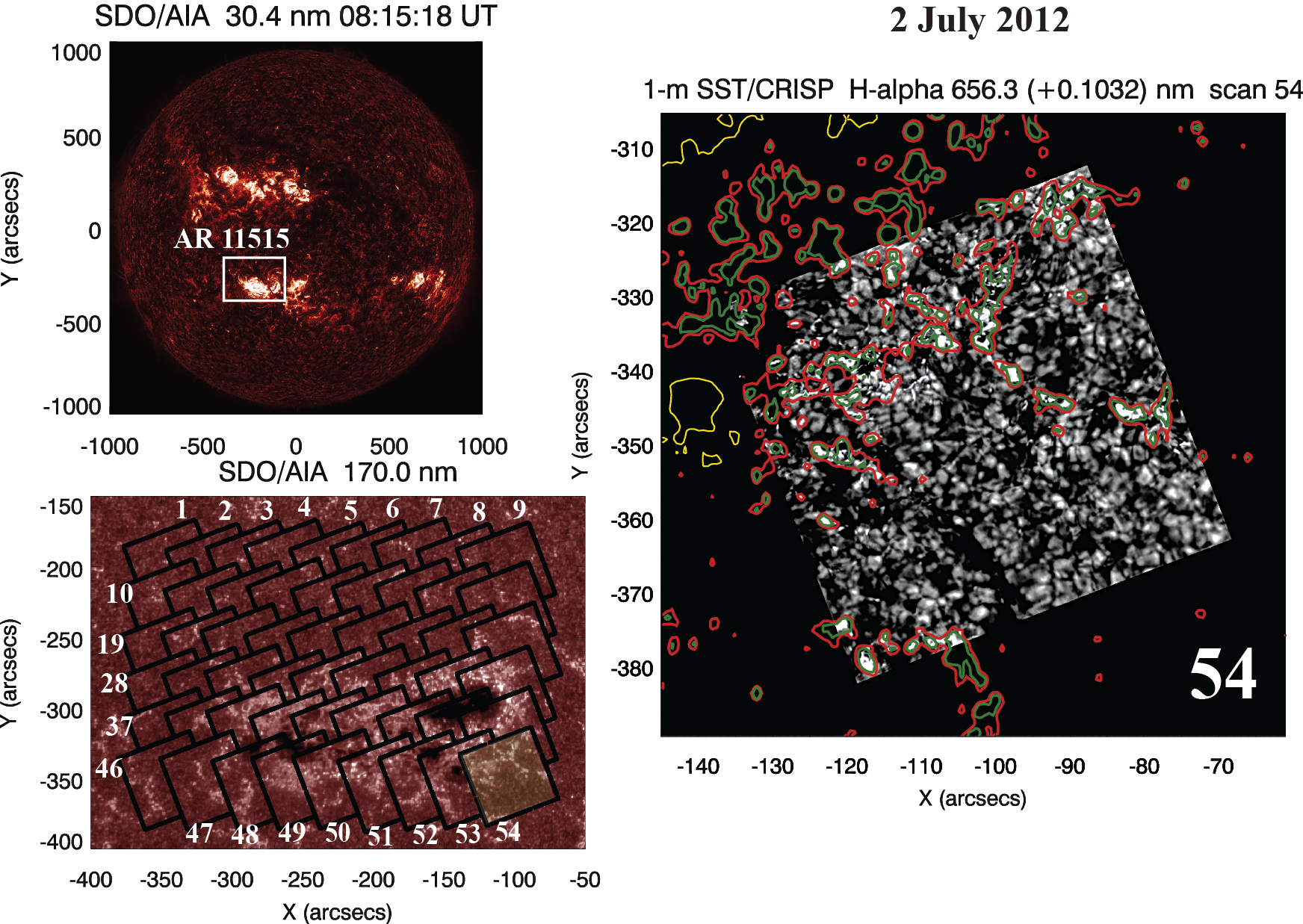}
\caption[The SST active region mosaic.]{An overview of the CRISP 54 grid mosaic of 2$^{nd}$~July AR 11515 is presented in the context of SDO/AIA 170.0~nm (lower left) and 30.4~nm (upper left: with extended FOV boxed) passbands. The order of the 5~min scan sequence (which was repeated once over a 10 min interval) is depicted (lower left), as a series of overlapping segments corrected for solar tilt. The accurate co-alignment of bright points in 170.0~nm (contoured in green and yellow), with coincident bright points in the grey-scale H-$\alpha$ continuum image from CRISP, is presented for grid segment no. 54 (right).}
\label{fig1}
\end{figure*}


\par However, in contrary to this \citet{2012ApJ...755L..33B} presented results of multi-stranded loop models calculated at a high resolution. They show that only five strands with a maximum radius of 280~km could reproduce the typical observed properties of coronal loops and a maximum of only eight strands where needed to reproduce all of the loops in their sample. More recently (11$^{th}$ July 2012), the High resolution Coronal Imager \citep[Hi-C: ][]{2013Natur.493..501C} took images of the 1.5~MK corona at a unprecedented resolution of 216---288~km \citep{2014ApJ...787L..10W}, which is unique for direct imaging of coronal loops in this passband. As a follow-up, \citet{2013ApJ...772L..19B} measured the Gaussian widths of 91 Hi-C loops observed in the solar corona and the resulting distribution had a peak width of 270~km. In other words, the finest-scale sub-structures of coronal loops are already observable. Other studies concerning the variations of intensity across a variety of hot loops, co-observed by AIA and Hi-C,  have continued to speculate on whether or not strand sub-structures could potentially exist well below what Hi-C or AIA can resolve \citep{2013A&A...556A.104P}. Most recently, \citet{2014ApJ...787L..10W} performed a statistical analysis on how the pixel intensity scales from AIA resolution to Hi-C resolution. They claim that 70\% of the Hi-C pixels show no evidence for sub-structuring, except in the moss regions within the FOV and in regions of sheared magnetic field. 


\begin{figure*}[!ht]
\includegraphics[scale=0.58, angle=0]{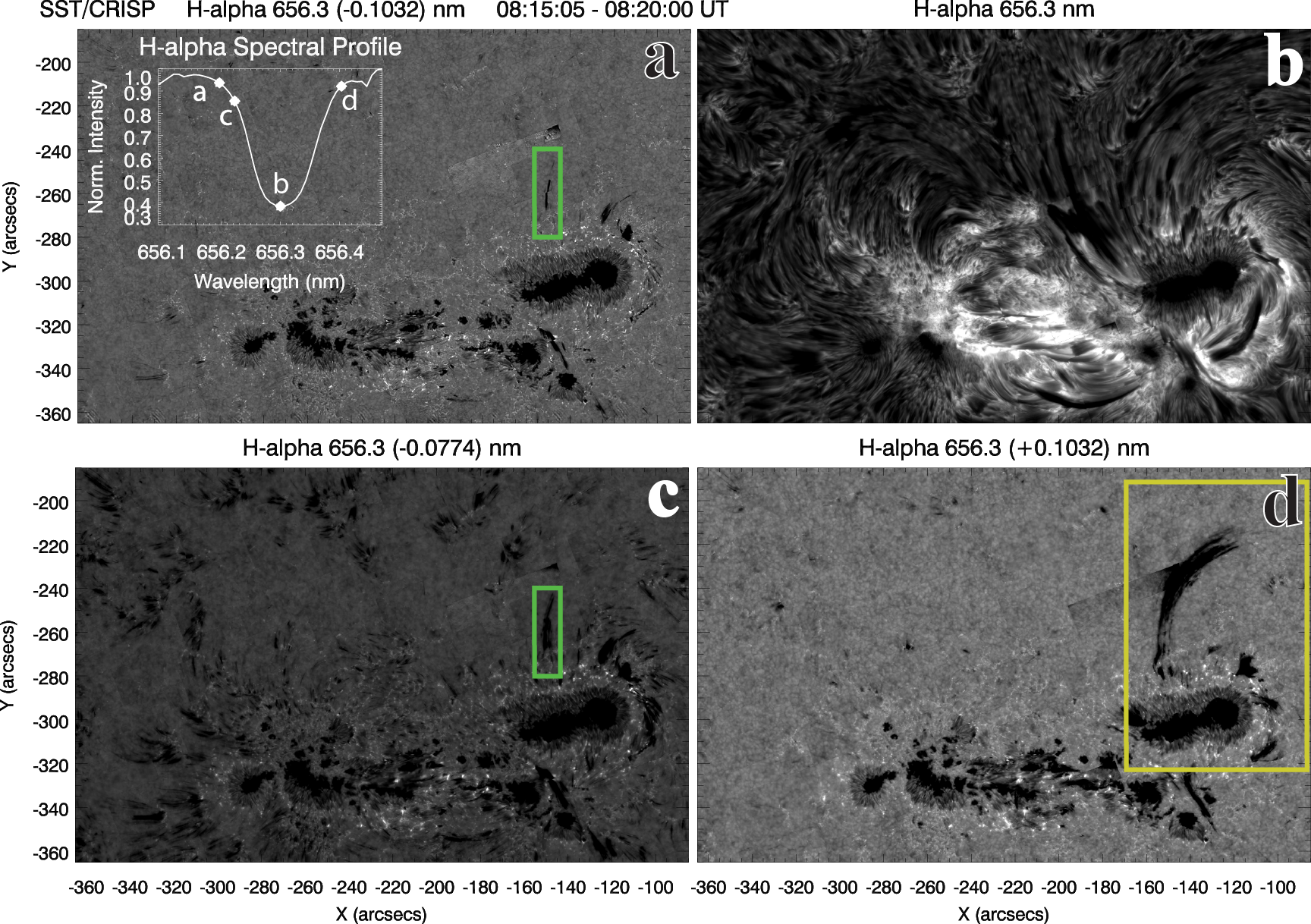}
\caption[The SST active region mosaic.]{CRISP H$\alpha$ line scan images for the reconstructed 54 grid mosaic of 2$^{nd}$~July, for the time interval 08:15:05 - 08:20:00~UT. The line scan includes two far wing positions (panels {\it a} in the far blue-wing and {\it d} in the far red-wing) . Panel {\it c} samples the fast spicular structures in the near blue-wing of H$\alpha$ and {\it b} samples the upper chromospheric plasma in the H$\alpha$ core, revealing a complex network of chromospheric loops and dynamic fibrils. The line scan positions, relative to the solar atlas H$\alpha$ profile, are presented in the sub-figure of panel ({\it a}) and the specific wavelengths of the scan are detailed in the panel titles. The green and yellow boxes mark features of interest for our investigation.}
\label{fig2}
\end{figure*}

%

\par There is strong evidence to suggest that coronal loops are, indeed, so finely structured when we consider loop legs from coordinated observations involving high resolution spectral imaging with ground-based instruments. \citet{2012ApJ...745..152A} performed a detailed and systematic study of coronal rain \citep{1970PASJ...22..405K, 2001SoPh..198..325S, 2004A&A...415.1141D} via the Swedish 1-m Solar Telescope \citep[SST: ][]{2003SPIE.4853..341S} / \citep[CRISP: ][]{2008ApJ...689L..69S} instrument at very high  spatial and spectral resolution (0\farcs0597 image scale). They detected narrow clumps of coronal rain in H$\alpha$ down to the diffraction limit (129~km) in cross-sectional area, with average lengths between $\sim$310~km and $\sim$710~km and widths approaching the diffraction limit of the instrument.  These measurements where repeated for on-disk coronal rain by \citep{2012SoPh..280..457A}. Coronal rain is considered to be a consequence of a loop-top thermal instability driving catastrophic cooling of dense plasma \citep[see ][and references therein]{2010ApJ...716..154A, 2012ApJ...745..152A}. Radiation cooling of dense evaporated plasma (filling coronal loops), leads to the onset the plasma depletion from the loops, slowly at first and then progressively faster. 


\begin{figure*}[!ht]
\centering
\includegraphics[clip=true,trim=0cm 0cm 0cm 0cm,scale=0.58, angle=0]{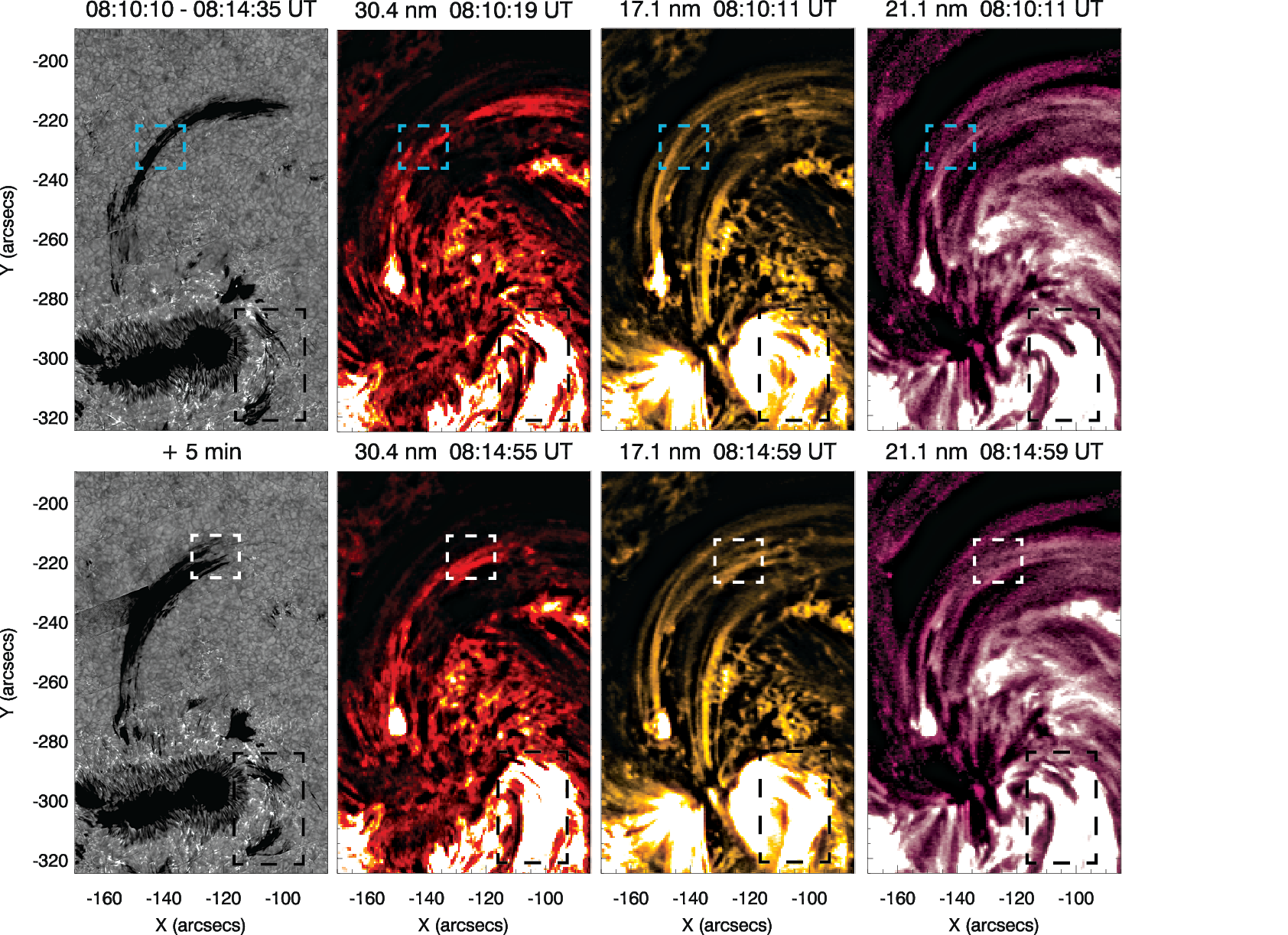}
\caption[The SST active region mosaic.]{The large-scale, high velocity down-flow in the H$\alpha$ far wing image ($+0.1032$~nm) is presented in the 1$^{st}$ panel ({\it top row}) and its evolution after 5~min, is presented in the 1$^{st}$ panel ({\it bottom row}). The co-spatial and co-temporal warm coronal loop is visible in the SDO/AIA He~{\sc ii}~30.4~nm (2$^{nd}$ column), Fe~{\sc ix}~17.1~nm (3$^{rd}$ column) and Fe~{\sc xiv}~21.1~nm (4$^{th}$ column) images as shown. The blue dashed box ({\it top row}) represents the FOV of Fig.~\ref{fig4a}, for a closer inspection of the coronal {\bf loop-leg} sub-structures. The white dashed box in the ({\it bottom row}) represents the coronal {\bf loop-top} and the FOV of Fig.~\ref{fig4b}.}
\label{fig3}
\end{figure*}


\par From an observational stand-point, the next step is to reveal evidence for sub-structuring along the full length of the coronal loop, from loop-top towards foot-point, and directly measure the threaded nature of coronal loop-tops using CRISP (with the most powerful resolving capability), in order to adequately test the existence of unresolved structure in the outer solar atmosphere. Observations from the ground-based instruments, such as CRISP, have obvious advantages over space-based facilities, with respect to resolving power given their much larger apertures. Analysis of the H$\alpha$ line core from such imaging spectro-polarimeters, provides an excellent tracer of the magnetic environment of the lower solar atmosphere \citep{2012ApJ...749..136L}. Through coordinated observations of coronal loops, with AIA and CRISP, we can use imaging in H$\alpha$, as a proxy for revealing the internal magnetic structure of coronal loops. As discussed in \citet{2012ApJ...745..152A}, it is possible that there is a strong dynamic coupling between neutrals and ions in coronal loops during the formation of coronal rain. As a result, the rain can become observable in the H$\alpha$ to reveal, in great detail, the topological structure of the local coronal magnetic field. The condensation process generates initially small rain clumps 'in situ' within coronal loop-tops, until the point where the mass density of the rain becomes large enough. leading to a flow of clumps towards the loop foot-points \citep{2013ApJ...771L..29F}. The lower limit (in spatial scales), with respect to the size distribution of these clumps, is dependent upon the magnetic fine-scale structuring of coronal loops. Momentarily and as a consequence of the thermal properties of the dense plasma undergoing rapid condensation, the H$\alpha$ signal is detectable within post-flare coronal loops, because the atmospheric conditions in the loops match that of the chromosphere. Fundamentally, the magnetic sub-structuring of the coronal loops near loop-tops, should remain the same or similar for all coronal loops (flaring and non-flaring). However, the possibility to probe the fine-scale magnetic structure of coronal loops, during the post-flare phase with H$\alpha$ via high resolution imaging, can present itself. 

\par In this letter, we report on the distribution of threaded sub-structures within a coronal loop from loop-top towards foot-point, via direct imaging of coronal loop cross-sections through a coordinated SST and SDO analysis, which comprises three datasets. 

\section{OBSERVATIONS}

\par CRISP, installed at the SST, is an imaging spectro-polarimeter that includes a dual Fabry-P{\' e}rot interferometer (FPI), as described by \citet{2006A&A...447.1111S}. The resulting FOV is about 55\arc$\times$55\arc. CRISP allows for fast wavelength tuning ($\sim$50~ms) within a spectral range and is ideally suited for spectroscopic imaging of the chromosphere. For H$\alpha$~656.3~nm the transmission FWHM of CRISP is 6.6 pm and the pre-filter is 0.49~nm. The image quality of the time series data is greatly benefited from the correction of atmospheric distortions by the SST adaptive optics system \citep{2003SPIE.4853..370S} in combination with the image restoration technique Multi-Object Multi-Frame Blind Deconvolution \citep[MOMFBD: ][]{2005SoPh..228..191V}. Although the observations suffered from seeing effects, every image is close to the theoretical diffraction limit for the SST. We refer to \citet{2008A&A...489..429V} and \citet{2013ApJ...769...44S} for more details on the MOMFBD processing strategies applied to the CRISP data. We followed the standard procedures in the reduction pipeline for CRISP data \citep{2014arXiv1406.0202D}, which includes the post-MOMFBD correction for differential stretching suggested by \citet{2012A&A...548A.114H}.


\begin{figure}[!h]
\includegraphics[clip=true,trim=0cm 0cm 0cm 0cm,scale=0.4, angle=0]{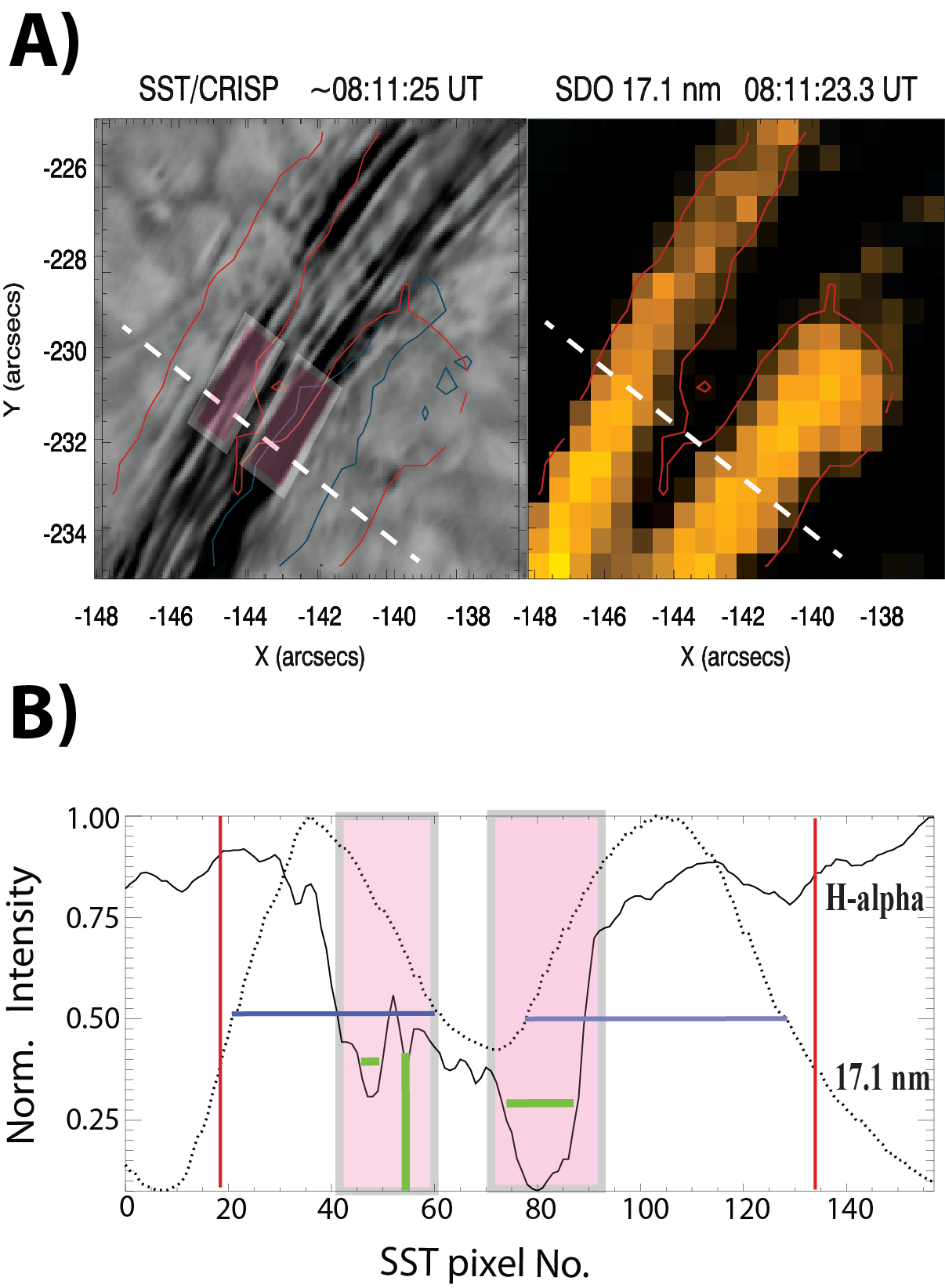}
\caption{Fine-scale, multi-stranded and multi-thermal sub-structures are detected within the coronal {\bf loop-leg} and presented here for the white-dashed box region.of Fig.~\ref{fig3}. The H$\alpha$ line position of +0.1032~nm image (grey-scale), is shown in panel-{\it A}, together with the near-simultaneous and co-spatial AIA 17.1~nm image. The coronal loop in 17.1~nm is contoured (solid red line) and overlaid in both images to compare with the H$\alpha$ multi-threaded component of the loop. A white-dashed diagonal slit and two pink-boxed regions are extracted and their normalized intensity profiles are plotted in panel-{\it B} for comparison of both spectral lines. The data cross-cuts for H$\alpha$ (solid curve) are overlaid with 17.1~nm (dotted curve). The blue solid lines presented the FWHM of the double peaked 17.1~nm profile and the green lines demarcate the locations of fine-scale strands which exist within the loop system.}
\label{fig4a}
\end{figure}


\par We explore the fully processed datasets with \citet[CRISPEX][]{2012ApJ...750...22V}, a versatile code for analysis of multi-dimensional data-cubes. We have compiled, with these reduction methods, three datasets from excellent periods of seeing which contained active region coronal loops within the CRISP FOV. 
\\
\par {\bf Dataset A} - A mosaic observing sequence (presented in Fig.~\ref{fig1} and Fig.~\ref{fig2}) was set to repeat once, with a line scan in H$\alpha$ at 4 wavelength positions (as presented in Fig.~\ref{fig2}), for the 280\arc~$\times$~180\arc~FOV containing Active region (AR) 11515 on 2$^{nd}$~July~2012, centred at [-225\arc,-275\arc] in solar-{\it x}/{\it y}. The pointing per position was preset to 5.5~s including 1.5-2~s for the telescope to change pointing. The total duration of the observation was 600~s so there were 108 pointing sequences in this interval with a repeat of 54 positions resulting in a cadence of 300~s between 08:10-08:20~UT. The co-alignment between CRISP and AIA for the mosaic observation is presented in Fig.~\ref{fig1}. 
\\
\par {\bf Dataset B} - A time series with 6 wavelength-point spectral scan in H$\alpha$ with an effective cadence of 19~seconds (after frame selection on the MOMFBD restored data) pointed at [-349\arc,-329\arc] in solar-{\it x}/{\it y} on 1$^{st}$ July 2012, centering on AR 11515 between 15:08-16:31~UT.
\\
\par {\bf Dataset C} - A time series with 43 wavelength-point spectral scan in H$\alpha$ with an effective cadence of 10.8~seconds pointed at [-818\arc,179\arc] in solar-{\it x}/{\it y} on 24$^{th}$ Sept. 2011, centering on AR 11302 between 10:17-11:02~UT.
\\
\par To achieve sub-AIA pixel accuracy in the co-alignment of H$\alpha$ with CRISP, and SDO/AIA, we cross-correlate photospheric bright points as observed in both instruments. Photospheric bright points exist as discrete, bright and relatively long lived features which exist in both quiet Sun, active regions and to a lesser extent in coronal holes. They are well distributed over the solar disk and can be clearly identified in the upper-photospheric AIA~170.0 nm channel (log T = 3.7). Our spectral line scan of H$\alpha$ includes nearby-continuum positions in both the blue wing ($-0.1032$~nm) and red wing ($+0.1032$~nm) for all our datasets. In the case of dataset {\bf A}, each grid in the mosaic sequence at the near-continuum spectral position was independently aligned to the corresponding SDO/AIA 170.0~nm (derotated FOV to compensate for solar rotation) in space and time. This accurate method for achieving a sub-AIA pixel accuracy in the co-alignment between the space-based SDO/AIA and ground-based SST/CRISP images is displayed in Fig.~\ref{fig1} (right). However, the same method was also applied in regard to the co-alignment of datasets {\bf B} and {\bf C}.

\section{RESULTS}

Dataset {\bf A}, composed of the mosaic coronal loop observation, contains primarily warm active region loops.. We examine the coronal loop multi-thermal sub-structures of active region loops in the following section. After that, we will focus on the sub-structures within loops in datasets {\bf B} and {\bf C}, which both contain hot, post-flare coronal loops. In that section, we will investigate the sub-structure of hot post-flare coronal loops which have experienced strong chromospheric evaporation.

\newpage

\subsection{ACTIVE REGION LOOPS}

\par In Fig.~\ref{fig2} we present the reduced and reconstructed mosaic images for the H$\alpha$ line scan of AR~11502 from the 2$^{nd}$~July 2012. The line scan positions include $-0.1032$~nm ({\it a}) and $-0.0774$~nm ({\it c}) in the blue wing, relative to the line core ({\it b}) and one position in the red at $+0.1032$~nm ({\it d}). The green box in Fig.~\ref{fig2}({\it a}) indicates the location of a high speed chromospheric upflow with a strong Doppler shift with an equivalent velocity of $\sim$47\kms. Interestingly, the trajectory of this high speed upflow, in the H$\alpha$ green box, coincides with the foot-point of a large-scale loop in apparent down-flow in the H$\alpha$ yellow box . The imposing loop-like structure, which extends above the sunspot group, is clearly observed in the red wing of H$\alpha$, as revealed in the yellow box in Fig.~\ref{fig2}({\it d}). This structure is co-spatial with EUV coronal loops as observed in SDO / AIA and its loop-top fine-scale structure is the focus of our investigation.


\begin{figure}[!hb]
\includegraphics[clip=true,trim=0cm 0cm 0cm 0cm,scale=0.4, angle=0]{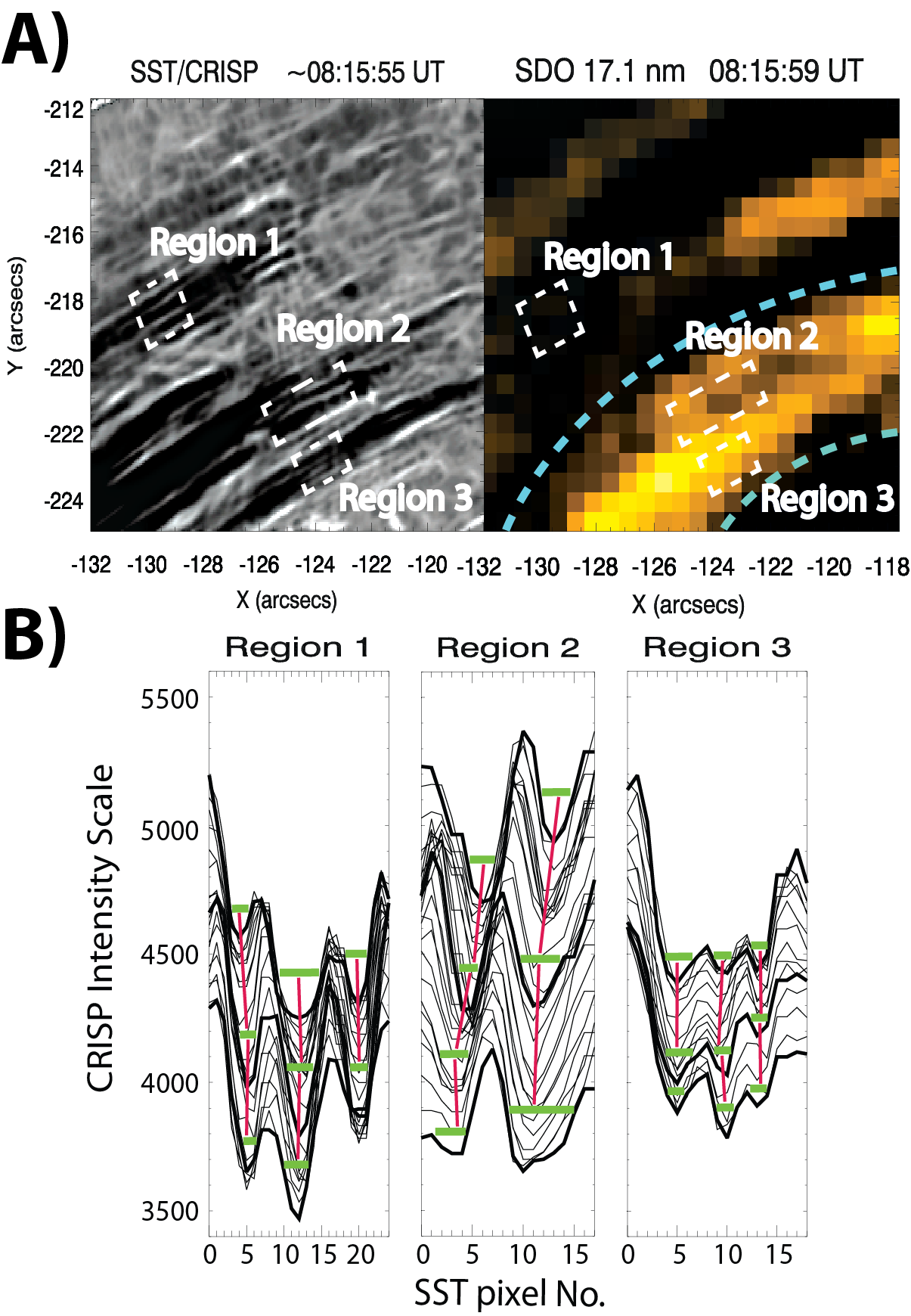}
\caption{Fine-scale, multi-stranded and multi-thermal sub-structures are detected within the coronal {\bf loop-top} and presented here for the blue-dashed box region.of Fig.~\ref{fig3}. The H$\alpha$ line position of +0.1032~nm image (grey-scale), is shown in panel-{\it A}, together with the near-simultaneous and co-spatial AIA 17.1~nm image. {\it Regions 1-3} in panel-{\it A} are selected for investigation of the H$\alpha$ intensity profile, as data cross-cuts along the loop-top system, which are represented in panel-{\it B}. As with Fig.~\ref{fig4a} parallel strands are identified using green lines separated by pink lines marking strand channels. {\it Regions 2} and {\it 3} are particularly highly structured in the H$\alpha$ line profiles.}
\label{fig4b}
\end{figure}


\par In Fig.~\ref{fig3}, we present a zoom into the yellow box (2) from Fig.~\ref{fig2}({\it d}) to reveal a multi-thermal sub-structure within the coronal loop system. The mosaic sequence was repeated with a 5-min time lag (before: {\it top row} and after: {\it bottom row}). When we compare between these time frames in the H$\alpha$ far red wing we can immediately reveal the evolution of the flow from the loop-top arching along the loop-leg. This evolution of the plasma is also evident in the AIA EUV lines at the loop-top in the same time interval, which confirms our expectation that the H$\alpha$ signature must originate within the coronal loop structure and at the loop-top. The loop length, as observed in H$\alpha$, is $\sim$63~Mm from lower loop leg to the central part of the loop-apex. Along the loop leg, near to the foot-point, a high velocity chromospheric up-flow (green box from Fig.~\ref{fig2}({\it a})) can be identified as a hot explosive event, which remains in emission in all AIA channels through the duration of the observation. This explosive event continues even after the excessive cooling of the loop-top and onset of the return flow to account for the high concentration of dense plasma near the loop-top. The blue-dashed boxes in the {\it top row} correspond to the FOV of the loop-leg which will be examined in more detail. Similarly, the pink-dashed boxes correspond to the FOV of the loop-top which we can investigate for signatures of fine-scale structure.

\par In Figs.~\ref{fig4a}~\&~\ref{fig4b} panel-{\it A}, we present the zoomed-in regions from Fig.~\ref{fig3} blue-dashed box (loop-leg).and white-dashed box (loop-top), respectively. Here we reveal in great detail, that both the loop-tops and loop-legs consist of bundles of fine-scale strand sub-structures, which can remain connected along the length of the loop. The fine strands, identifiable within the pink boxed regions in Fig.~\ref{fig4a} panel-{\it A} and white-dashed in Fig.~\ref{fig4b} panel-{\it A}, appear to be parallel with each other and exist / contained within the AIA-defined loop boundary (see the blue dashed line in Fig.~\ref{fig4b} panel-{\it A}). In Fig.~\ref{fig4a} panel-{\it B}, the data cross-cuts (extracted from the diagonal-dashed slit of Fig.~\ref{fig4a} panel-{\it A}) demonstrate the sub-structured nature of the loop in H$\alpha$ within, and not necessarily confined too, the double peak profile representing 17.1~nm normalised intensity. 


\begin{figure}[!h]
\includegraphics[scale=0.4, angle=0]{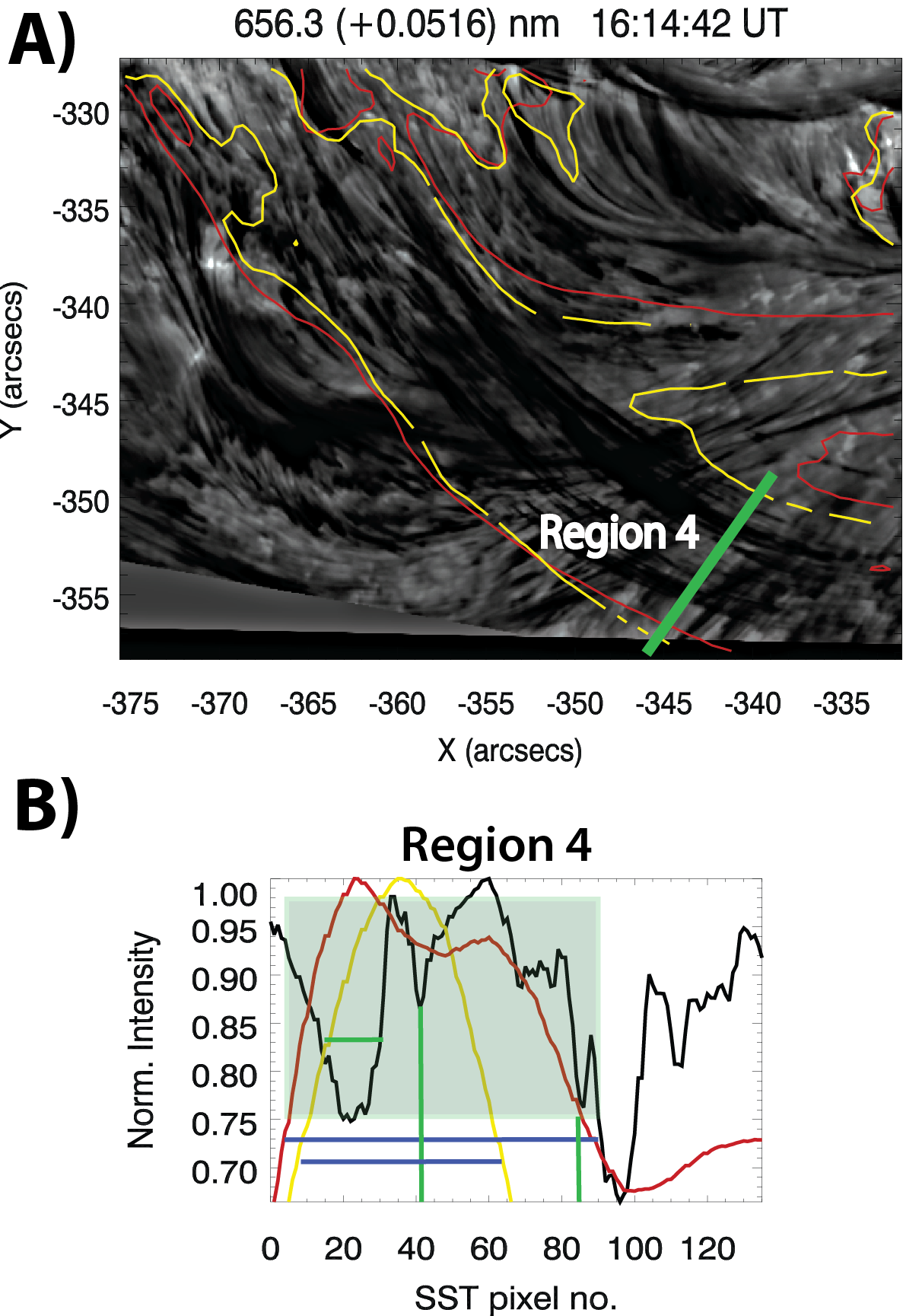}
\caption{Co-temporal and co-spatial H$\alpha$ near red-wing images (grey-scaled), together with overlaid contours (17.1~nm: yellow and 21.1~nm: red), are presented in panel-{\it A}. The observations consists of a snapshot of a post C8.2-class flare system from 1$^{st}$~July~2012 (dataset {\bf B}). Panel-{\it B} presents the normalised intensity cross-cuts of the post-flare {\bf loop-top} (solid green line {\it Region 4} in panel-{\it A}) for the associated H$\alpha$ signal (black curve) along with the respective curves of the 17.1~nm (yellow) and 21.1~nm (red) channels, as is contoured in panel-{\it A}. The shaded green-boxes represent examples of associated fine-scale structures in H$\alpha$ and the EUV lines from which we extract measurable strand cross-sections for our statistical sample. The blue-horizontal lines represent the well-defined and measurable cross-sections of the EUV loops in contrast with the fine-scale structuring in H$\alpha$.}
\label{fig5a}
\end{figure}


The 17.1~nm loop system boundary (marked by the vertical red lines in Fig.~\ref{fig4a} panel-{\it B}) has a maximum cross-sectional width of 3-4 Mm and also appears structured down to the resolution limit of the AIA instrument. The AIA temperature response function for the 17.1~nm passband has its maximum around 0.9~MK.  We find that the Full-Width-Half-Maximum (FWHM) of each of the AIA 17.1~nm loop peaks, from the cross-cut, is 870~km. In measuring the cross-sectional widths, we computed the FWHM of the data cross-cut as being the width of the bisector corresponding to half of the difference between the minimum (in the case of H$\alpha$) or maximum (in the case of 17.1~nm) intensity level and the background pixel intensity level.  In Fig.~\ref{fig4a} panel-{\it B}, at the centre of this AIA loop double peak, we can identify a very finely structured bunch of strands in H$\alpha$. The H$\alpha$ data cross-cuts contain multiple, parallel strands with a variety of cross-sectional widths (individual strands are marked with the green solid lines). The broadest strand in this set has a FWHM of 6 SST-pixels which corresponds to 258~km. This relatively broad strand extends with a consistently uniform cross-section to a length of 4855~km. The left-most strand is narrower again corresponding to 4 SST-pixels which is 172~km. Other strands appear to exist within a range of spatial scales that can be as large as 516~km and as low as 129~km. On average, we can detect a maximum of 8 strands within this cross-section of the loop. 

\par Similarly, in Fig.~\ref{fig4b} panel-{\it B}, H$\alpha$ data cross-cuts are plotted in sequence depicting the strands parallel channels (see the connecting green and pink solid lines demarcating the channel of the strand), that run along the length of the loop within the {\it Regions 1, 2} and {\it 3}. {\it Regions 2} and {\it 3} are contained within, and bounded by, the curved blue-dashed lines of the AIA 17.1~nm loop boundary. This is the same loop system connecting the loop-leg from Fig.~\ref{fig4a}. The finest detectable strands exists within {\it Region 3} and they all extend uniformly in length, across the FOV for at least 1100~km. The right-most strand channel from {\it Region 2} has a cross-sectional width similar to that of the broadest strand from the loop-leg section. However, the most commonly occurring strand cross-section, from both loop-top and loop-leg sections, is 129~km. 


\begin{figure}[!h]
\includegraphics[scale=0.4, angle=0]{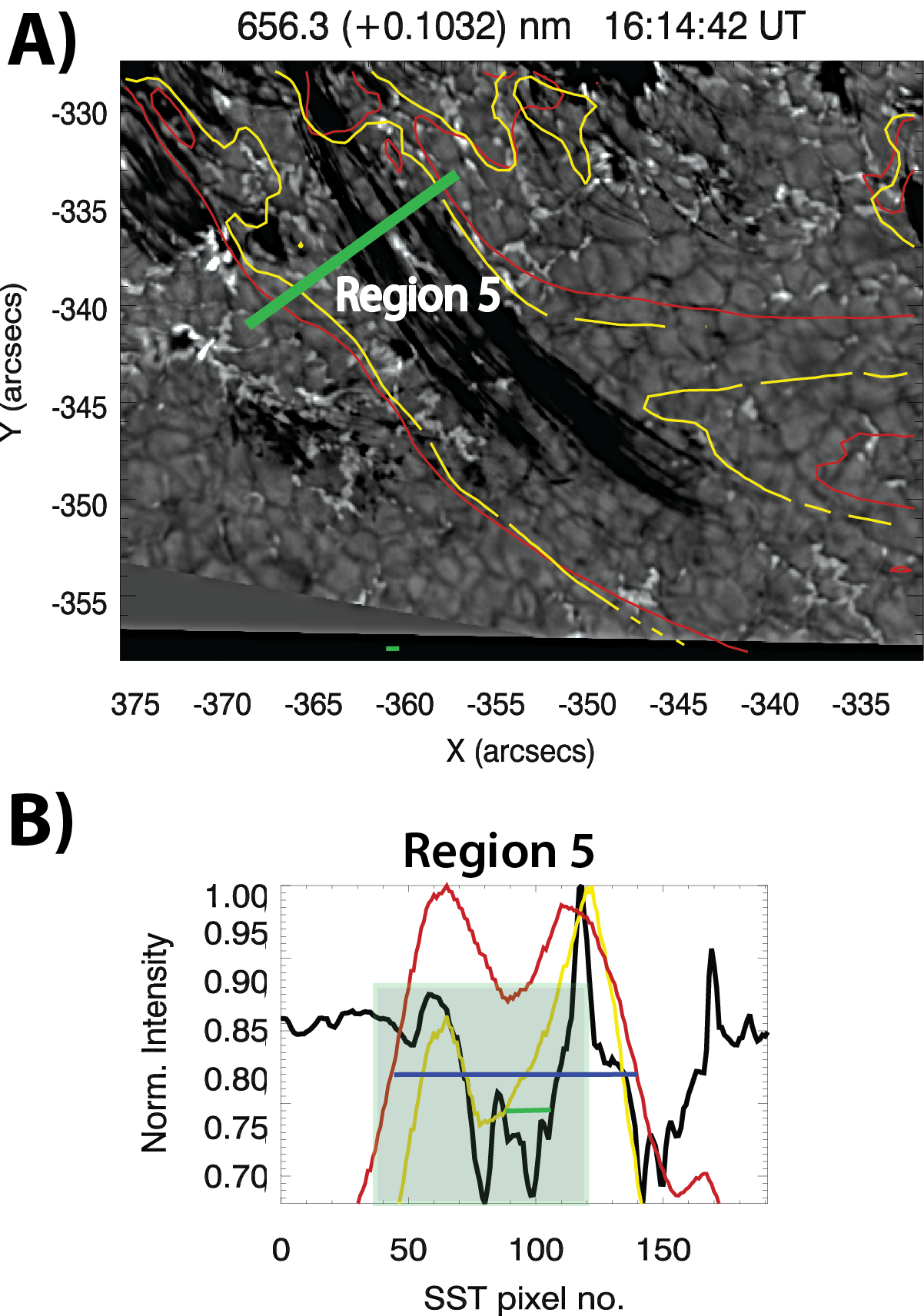}
\caption{Co-temporal and co-spatial H$\alpha$ far red-wing images (grey-scaled), together with overlaid contours (17.1~nm: yellow and 21.1~nm: red), are presented in panel-{\it A}. The observations consists of a snapshot of a post C8.2-class flare system from 1$^{st}$~July~2012 (dataset {\bf B}). Panel-{\it B} presents the normalised intensity cross-cuts of the post-flare {\bf loop-leg} (solid green line {\it Region 5} in panel-{\it A}) for the associated H$\alpha$ signal (black curve) along with the respective curves of the 17.1~nm (yellow) and 21.1~nm (red) channels, as is contoured in panel-{\it A}. The additional markers in these figures are previously described in Fig.~\ref{fig5a} for this dataset.}
\label{fig5b}
\end{figure}


\par  We have found multi-stranded, and multi-thermal fine-scale structuring within warm active region coronal loops in both loop-top and loop-leg sections. However, is this scenario consistent with hotter post-flare loops, which can undergo a more widespread and intense foot-point heating? 

\subsection{POST-FLARE LOOPS}

\par Figs.~\ref{fig5a}~\&~\ref{fig5b} display the overlays of the hot post-flare loop system of dataset {\bf B}, which consists of CRISP observations centred on AR~11515 (same active region as dataset {\bf A}, observed one day later) on the 1$^{st}$~July~2012 and hosts a C8.2-class flare during the observation period. In both Figs.~\ref{fig5a}~\&~\ref{fig5b} panel {\it A}, the post-flare loops are presented 33~mins after the flare peaks at $\sim$15:41 UT in the GOES X-ray channel. The loop boundaries are contoured in 21.1~nm (red) and 17.1~nm (yellow) channels and overlaid on the H$\alpha$ red wing images. In Fig.~\ref{fig5a} panel-{\it A} the H$\alpha$ spectral line position is +0.0516~nm whereas in Fig.~\ref{fig5b} it is further into red wing (at  +0.1032~nm) where we detect the faster moving components within the post-flare loop. Its is immediately obvious that in panel-{\it A} for both Fig.~\ref{fig5a}~\&~\ref{fig5b}, we detect a clear spatial correlation between the cooler fine-scale structures in H$\alpha$ and the hotter EUV signal. The condensing coronal rain (known to form during the post-flare cooling phase) is shown to be depleting from the loop-top (located near {\it Region 4}) and travels towards the loop foot-point. Using the spectral scans in H$\alpha$ we can sample the fine-scale structure within the cross-section of the multi-thermal loop system, and with very high accuracy since the EUV loops are very intense therefore very well defined in the images. The cross-cut data for {\it Regions 4} and {\it 5}, from Fig.~\ref{fig5a}~\&~\ref{fig5b} panel-{\it B}, displays similar fine-scale structuring very much confined within the post-flare loops. Again we detect double peak structures in many of the AIA channels (notably in the 21.1~nm red curves) and multiple strands in the H$\alpha$ line position, indicating the presence of narrower threads well below the instrumental resolution of AIA. The green-boxed regions in Fig.~\ref{fig5a}~\&~\ref{fig5b} panel-{\it B}, highlight the sections of the cross-cuts where we have an overlap between AIA loops and structure in H$\alpha$, which again implies a coincidence in the location of the formation of the lines to within the coronal loop itself. The cross-sectional width measurements of all the fine-scale strands, within the full length of the loop system from loop-top to foot-point, will be accumulated together with datasets {\bf A}, for statistical comparison of the variation in the range of scales present within the loop systems.


\begin{figure}[!hb]
\includegraphics[scale=0.4, angle=0]{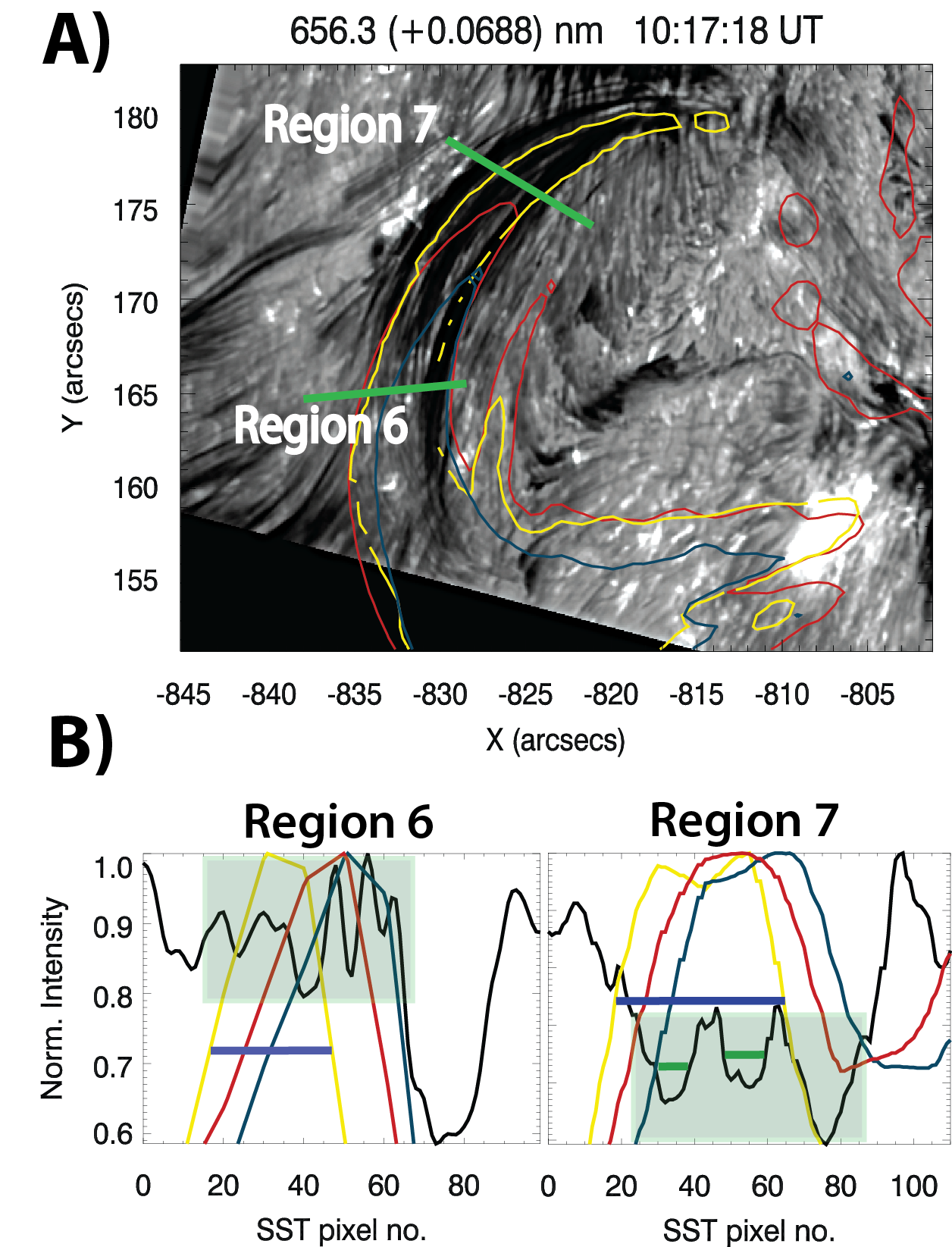}
\caption{Co-temporal and co-spatial H$\alpha$ red-wing images (grey-scaled), together with overlaid contours (17.1~nm: yellow and 21.1~nm: red), are presented in panel-{\it A}. The observations consists of a snapshot of a post X1.9-class flare loop system from 24$^{th}$~Sept.~2011 (dataset {\bf C}). Panel-{\it B} presents the normalised intensity cross-cuts of the post-flare loop-leg (solid green line {\it Region 7} in panel-{\it A}) for the associated H$\alpha$ signal (black curve) along with the respective curves of the 17.1~nm (yellow) and 21.1~nm (red) channels, also contoured in panel-{\it A}. Likewise, we also plot intensity profiles for {\it Region 6}, representing fine-scale structure close to the loop-top, in panel-{\it B}. The shaded green-boxes represent examples of associated fine-scale structures in H$\alpha$ and the EUV lines from which we extract measurable strand cross-sections for our statistical sample. The blue-horizontal lines represent the well-defined and measurable cross-sections of the EUV loops in contrast with the fine-scale structuring in H$\alpha$.}
\label{fig6a}
\end{figure}


\par
\begin{figure}[!h]
\includegraphics[scale=0.4, angle=0]{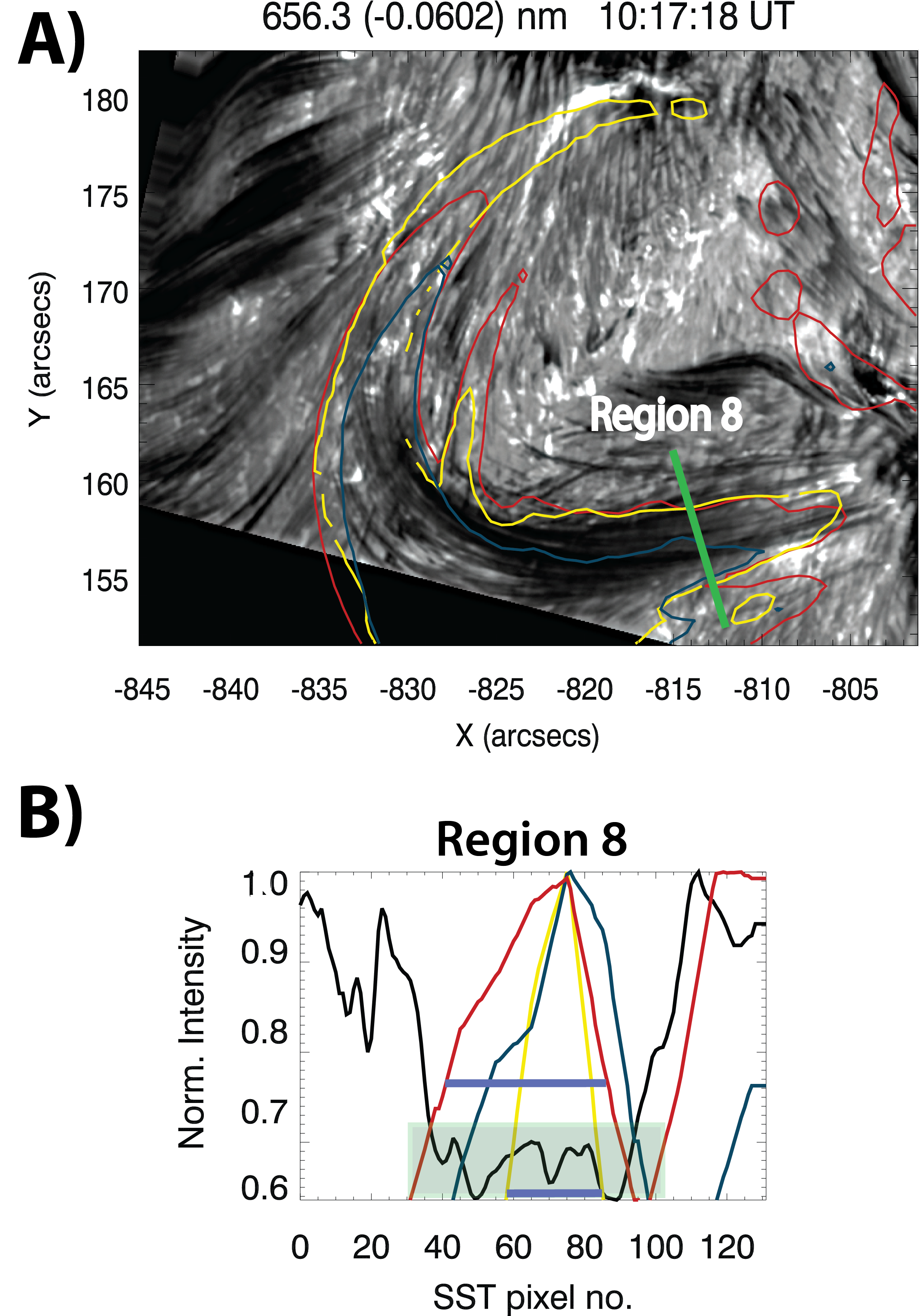}
\caption{Co-temporal and co-spatial H$\alpha$ blue-wing images (grey-scaled), together with overlaid contours (17.1~nm: yellow and 21.1~nm: red), are presented in panel-{\it A}. The observations consists of a snapshot of a post X1.9-class flare system from 24$^{th}$~Sept.~2011 (dataset {\bf B}). Panel-{\it B} presents the normalised intensity cross-cuts of the other post-flare {\bf loop-leg} (solid green line {\it Region 8} in panel-{\it A}) for the associated H$\alpha$ signal (black curve) along with the respective curves of the 17.1~nm (yellow) and 21.1~nm (red) channels, as is contoured in panel-{\it A}. The additional markers in these figures are previously described in Fig.~\ref{fig6a} for this dataset.}
\label{fig6b}
\end{figure}

 Dataset {\bf C} consists of CRISP observations centred on a region that hosted a GOES X1.9-class flare (post-impulsive phase close to the north-east solar limb) on the 24$^{th}$~Sept.~2011. In Figs.~\ref{fig6a}~\&~\ref{fig6b} panel-{\it A}, we present images of the H$\alpha$ red-wing (Fig.~\ref{fig6a}) and blue-wing (Fig.~\ref{fig6b}) grey-scaled images, which are almost coincident in observation time at 10:17~UT (within the same line scan with one line scan taking 4.2 seconds). These images are again overlaid with contours from AIA 17.1~nm (yellow), 21.1~nm (red) and also 19.3~nm (dark blue), for the post flare loop system 56~mins after the flare GOES X-ray peak. During this phase we again expect to see evidence of coronal rain formation, with a characteristic signature in H$\alpha$. In dataset~{\bf C}, catastrophic cooling is indeed present across the entire post-flare loop and the opposing flows, as the rapidly cooled coronal rain depletes from the loop-top under the action of gravity, is exquisitely revealed. We can clearly detect opposing, dark Doppler flows running along both legs of the loop system, as absorption in H$\alpha$ (as defined by AIA contours), in the red-wing for the left-side leg (Fig.~\ref{fig6a}) and in the blue-wing for the right-side leg (Fig.~\ref{fig6b}). The Doppler signature in the loop reveals the geometrical nature of the loop itself. In the line core images we can detect the structure of the loop-top where there is a net zero Doppler shift, which is in agreement with the location of the observed loop-top in the EUV channels and confirmation that the hot and cold plasma must be co-located within the same loop structure. 


As with the large-scale loop of dataset~{\bf A}, in the H$\alpha$ line core images of dataset~{\bf C}, many of these finely-structured strands extend from foot-point to foot-point (crossing the loop-top). As with dataset~{\bf B}, in Fig.~\ref{fig6a}~\&~\ref{fig6b} panel-{\it B}, we present sets of normalised intensity cross-cuts from both H$\alpha$ (black curve) and the EUV channels (17.1~nm: yellow and 21.1~nm: red curves) for slit {\it Regions 6, 7} and {\it 8}.  Again, we can reveal fine-scale structuring within the H$\alpha$ intensity profiles (from within the green shaded boxes), in both Fig.~\ref{fig6a}~\&~\ref{fig6b}, that are co-spatial with singly-peaked profiles in the EUV channels. In each of the {\it Regions 7} and {\it 8}, we can detect similarly scaled strands and a variable range in the cross-section strand number density of, typically, 3-5 clearly defined parallel strands. The data cross-cuts from each of the AIA passbands contoured here are also overlaid. The EUV loop intensity profiles appear to have cross-sectional widths in the range of 25-40 SST pixels (approx. 5-6 times greater than fine-scale H$\alpha$ strands within them as marked by solid green lines), as marked with the solid blue lines, in panel-{\it B} of both Fig.~\ref{fig6a}~\&~\ref{fig6b}. These measurements are comparable with those profiles deduced from dataset~{\bf B} which was a substantially weaker post-flare loop system, and likewise, for dataset~{\bf A} for a warm active region loop system with indication of foot-point heating but no apparent flaring.

\par A statistical comparison of the strands, using all of the examples from each of the datasets examined, is considered next. In summary, these coronal loop sub-structure samples comprise both loop-top and loop-leg sub-structures from both CRISP and AIA images, where we have a detectable and confident correspondence between H$\alpha$ features in CRISP and associated EUV loops in AIA. A total of 62 coronal loop sub-structures where measured from all three datasets. A resulting histogram of the number density of all the strands / loop cross-sections versus their cross-sectional widths (km units), is presented in Fig.~\ref{fig19}. Here we can show that the distribution of sub-structures within coronal loops appears to increase exponentially towards finer scales and the highest number density (10 instances representing almost a 6$^{th}$ of all sub-structures measured) appears to peak within the range of the CRISP resolution. In fact, the distribution would imply that we have not yet reached a peak in the dominant spatial fine-scale structure of coronal loops and furthermore, we are not yet within the observable range of the finest scales of structure.

\section{DISCUSSION AND CONCLUSIONS}


\begin{figure*}[!ht]
\centering
\includegraphics[scale=0.45, angle=0]{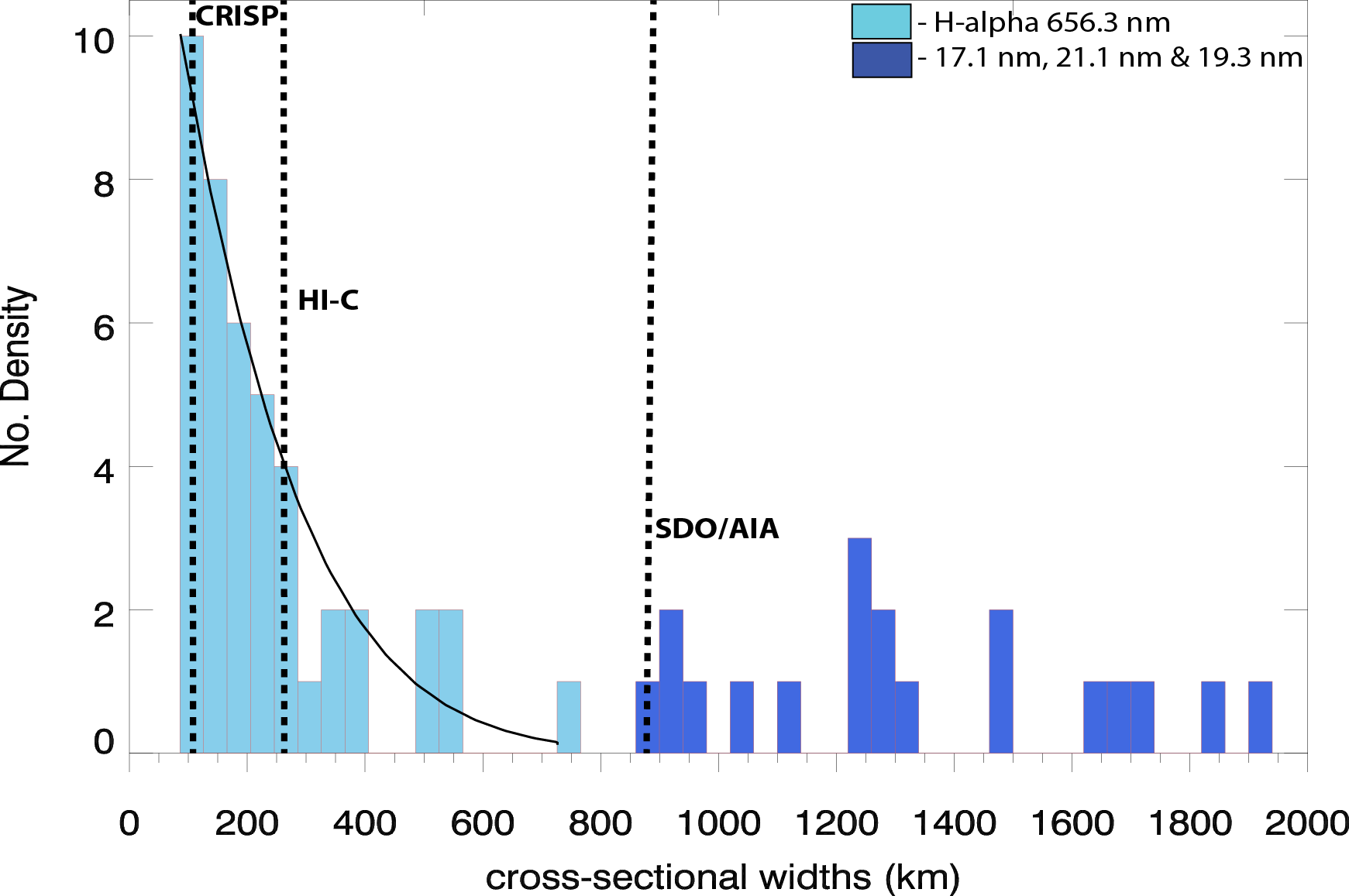}
\caption[The SST active region mosaic.]{The histogram displays the distribution of all detectable strands and sub-structures within coincident H$\alpha$ and EUV coronal loops, as measured from all of the datasets sampled. The pale blue sections correspond to the H$\alpha$ only detections made via CRISP. The darker blue sections correspond to the SDO / AIA coronal loop cross-sections. The vertical dashed lines mark the resolution limit for CRISP, Hi-C and SDO / AIA. The number density of detected strands versus their cross-sectional FWHM widths is measured. The exponential curve is overlaid onto the plot to indicate the steeping distribution towards finer scales within the sub-structures of coronal loops.}
\label{fig19}
\end{figure*}


\par Since the launch of Hi-C in 2012 there has been substantiative research into the fine-scale structure of coronal loops. Most efforts to address this issue are centred on multi-instrumental approaches / analysis, comparing statistical relationships of intensity variations between measured loop cross-sections, in co-incident Hi-C and AIA coronal loops. Ultimately, the conclusions from such studies, as with this study, are always going to be limited by the resolution of the instruments used and any conclusions on the existence of fine-scale structure will continue to be speculated upon, until the necessary improvements in instrumentation resolving power are met. In this study, we take this investigation further by exploiting the resolving power potential of the ground-based CRISP instrument together with SDO / AIA coronal loop detections to reveal the fine-scale structure. 
 

\par Our analysis of three datasets, which consist of large-scale coronal loops in various conditions (ranging from warm active region loops to hot post-flare loops), have been accurately co-aligned with very high resolution imaging in H$\alpha$. Interestingly, there is little difference in the distribution of strand / structure spatial scales that would lead one to be able to distinguish between datasets~{\bf A}, {\bf B} and {\bf C}, each of which depict loop-systems undergoing large variations in impulsive heating. This aspect may be hinting that the magnetic sub-structure of coronal loop cross-sections may not be so sensitive to variations in loop foot-point heating or, alternatively, the magnetic field is effectively and systematically reacting to changes in the thermal properties of the internal loop environment, in order to manage the heat transport and maintain stability. The formation of the coronal rain is a demonstration of the loop system reaching a new thermal equilibrium, as observed in the cool H$\alpha$ line, which acts as a tracer of  the magnetic environment. This association of this rapid cooling condensation process and its temporary association with the EUV coronal loops has been exploited in this study, in order to examine the fine-scale structure of the loops at the loop apex. In Fig.~\ref{fig3}, we clearly demonstrate the loop-top depletion due to catastrophic cooling of plasma, which falls back to the lower solar atmosphere along the loop-leg. The longest continuous detectable strand (which largely features close to the resolution limit in CRISP at 129 km) was on the order of 26,100 km, extending from loop-top to close to the foot-point. This represents one of the longest and continuous fine-scale coronal loop substructures detected to date. 

\par \citet{2010ApJ...716..154A} observed coronal rain near coronal loop-tops with Hinode \citep[SOT: ][]{2008SoPh..249..167T} and measured cross-sectional widths on the order of 500~km. We detect similarly scaled coronal rain strands in coronal loop-tops with CRISP and also threads with finer scales, implying the existence of a range of finely scaled structures in the outer solar atmosphere. The draining of the dense plasma as it falls back towards the loop foot-points from the loop-top is most clear in dataset {\bf C}, represented in Figs.~\ref{fig6a}~\&~\ref{fig6b}. In the images, we demonstrate a clear association of the rain flowing within both legs of a post-flare coronal loop from its apparent source near the loop-top. There appears to be a distribution of scales within the coronal loop-top with respect to cross-sectional widths of strands. Likewise, there is a distribution in the strand lengths, all of which appear to follow the trajectory of the loop-top coronal field (as inferred from co-incident AIA loop trajectories), with some appearing to be very much extended towards the loop foot-point. This shows that the fine-scale structure is widespread along the full length of the loop and the coronal rain clumps can form within bunches of strands. We can conclude that the vast majority of fine-scale strand structures within coronal loop cross-sections exist well below the resolution of SDO /AIA (69.3\% of the potential strands, as returned by CRISP, are unresolved with AIA) and almost 50\% of fine-scale strands could potentially remain unresolved with imaging in an instrumental-resolution comparable to Hi-C.  In summary, after considering 8 cross-cuts (representing one loop-top and two loop-legs for datasets {\bf B} and {\bf C} and one loop-top, one loop-leg for dataset {\bf A}) we find an average ratio of 5 : 2 for CRISP strand no. density to AIA strand no. density per loop system.

\par Finally, we conclude that there is a cut-off in the peak of the distribution (from Fig.~\ref{fig19}), at the instrumental resolution of CRISP. We and others have assumed that the distribution of strand sizes should be a Gaussian or at least symmetric about some peak. From our histogram, we demonstrate that either we have not yet reached that peak and the actual fine-scale resolution is much below 100~km, or the spatial-scale distribution is in fact skewed away from being symmetric about some peak. This result clearly states that, even with the most powerful ground-based instrumentation available, we have not yet observed a true peak in the strand cross-sectional width measurements at the lowest limit within coronal loops. \citet{2013ApJ...771L..29F} demonstrated with numerical simulations of coronal rain formation that, when compared with observational statistics, a higher percentage of coronal rain clumps are expected in smaller scales. Here, we can confidently state that the peak (in other words the minimum) in cross-sectional width distribution of the finest structures within coronal loops, is most likely to exist beneath the 100~km mark. Henceforth, we look forward with great anticipation to the arrival of more powerful ground-based telescope facilities \citep[such as the 4-m Daniel K. Inouye Solar Telescope (DKIST): ][]{2013SPD....4440002B}, in order to fully probe even finer scales within coronal loop cross-sections. 


\acknowledgements

We further thank N~Frej who co-observed the mosaic and 1$^{st}$~July~2012 datasets ({\bf A} \& {\bf B}), as well as, A.~O.~Carbonell and B.~Hole who observed at the SST on the 24$^{th}$~Sept.~2011 dataset ({\bf C}). The Swedish 1-m Solar Telescope is operated on the island of La Palma by the Institute for Solar Physics of Stockholm University in the Spanish Observatorio del Roque de los Muchachos of the Instituto de Astrofisica de Canarias.

\bibliography{bib}

\bibliographystyle{aa}


 
\end{document}